\documentclass[numberedappendix]{emulateapj}

\usepackage{graphicx}
\newcommand{\degree}{^\circ}

\newcommand{\be}{\begin{equation}}
\newcommand{\ee}{\end{equation}}
\newcommand{\bp}{\begin{figure}[!ht]}
\newcommand{\ep}{\end{figure}}
\newcommand{\bpm}{\begin{figure*}[!ht]}
\newcommand{\epm}{\end{figure*}}
\newcommand{\reffig}[1]{Figure \ref{fig:#1}}
\newcommand{\refsec}[1]{\S \ref{sec:#1}}

\begin{document}




\title{Identifying the Radio Bubble Nature of the Microwave Haze}

\author{Gregory Dobler\altaffilmark{1,2}}

\altaffiltext{1}{
  Kavli Institute for Theoretical Physics,
  University of California, Santa Barbara
  Kohn Hall, Santa Barbara, CA 93106 USA
}
\altaffiltext{2}{
  dobler@kitp.ucsb.edu
}


\begin{abstract}
  Using 7-year data from the \emph{Wilkinson Microwave Anisotropy Probe}
I identify a sharp ``edge'' in the microwave haze at high Galactic
latitude ($35\degree < |b| < 55 \degree$) that is spatially coincident
with the edge of the ``\emph{Fermi} Haze/Bubbles''.  This finding
proves conclusively that the edge in the gamma-rays is real (and not a
processing artifact), demonstrates explicitly that the microwave haze
and the gamma-ray bubbles are indeed the same structure observed at
multiple wavelengths, and strongly supports the interpretation of the
microwave haze as a separate component of Galactic synchrotron (likely
generated by a transient event) as opposed to a simple variation of
the spectral index of disk synchrotron.  In addition, combining these
data sets allows for the first determination of the magnetic field
within a radio bubble using microwaves and gamma-rays by taking
advantage of the fact that the inverse Compton gamma-rays are
primarily generated by scattering of CMB photons at these latitudes,
thus minimizing uncertainty in the target radiation field.  Assuming
uniform volume emissivity, I find that the magnetic field within our
Galactic microwave/gamma-ray bubbles is $\sim5$ $\mu$G above 6 kpc off
of the Galactic plane.

\end{abstract}


\section{Introduction}
\label{sec:introduction}
  Recent full sky data sets by the \emph{Wilkinson Microwave Anisotropy
Probe} (WMAP) and the \emph{Fermi Gamma-Ray Space Telescope} have
revealed the presence of a new and very large structure in the Milky
Way.  This emission manifests as an excess of both microwaves and
gamma-rays when removing Galactic diffuse emission associated with
known emission mechanisms at these wavelengths and has come to be
named the ``WMAP Haze'' \citep{finkbeiner04a,dobler08a,dobler12} and
the ``Fermi Haze/Bubbles'' \citep{dobler10,su10} in the microwaves and
gamma-rays respectively.

In microwaves, the haze is synchrotron emission with a brightness
temperature as a function of frequency $T \propto \nu^{\beta_H}$ with
$\beta_H \approx -2.5$ \citep{dobler12}.  This spectral dependence
implies that the underlying electron spectrum (number density as a
function of energy) is given by $dN/dE \propto E^{\gamma}$ with
$\gamma \approx -2$ for energies $\sim$10 GeV.  In gamma-rays, the
haze/bubbles is most likely due to inverse Compton (IC) scattering of
the starlight, infrared, and cosmic microwave background (CMB)
interstellar radiation field (ISRF).\footnote{There is also the
possibility that the gamma-ray emission is due to the decay of $\pi^0$
particles generated by proton-proton collisions within the
haze/bubbles.  However, to match the amplitude and brightness profile,
this scenario relies on a $\sim10^9$ yr wind \citep[see][]{crocker11}
and it is difficult to reconcile the sharp edge with this long
timescale.  In addition, there is no associated H$\alpha$ signal as is
typically seen in winds and the spectrum appears consistent with an
inverse Compton scenario (see \refsec{results}).}  In the original
discovery paper of the \emph{Fermi} haze/bubbles,
\cite{dobler10} showed that the spectrum of the
gamma-rays is consistent with IC emission from the \emph{same electron
population responsible for the microwave haze synchrotron} in both
energy dependence as well as overall normalization.  However, as shown
by \cite{dobler12}, the existence of $\sim$1-10 GeV IC gamma-rays at
high latitude where the ISRF is dominated by the CMB require electrons
with energies $\sim$TeV.

Taken together, the microwaves and gammas imply that the electron
spectrum is roughly a powerlaw from $\sim$1-1000 GeV suggesting that
either the electrons have not had sufficient time to cool \citep[the
cooling time in this energy range is $\sim10^6$-$10^7$ yr;
see][]{su10} or are continuously being accelerated within the
haze/bubbles \citep[e.g.,][]{dobler11a,mertsch11}.  The large volume
of hard spectrum cosmic-rays has made it difficult to identify an
underlying origin for the haze/bubbles and there have been numerous
studies exploring the possibilities from
starbursts \citep{biermann10}, to Galactic winds \citep{crocker11}, to
jet blown bubbles \citep{guo11a,guo11b}, and even co-annihilation of
dark matter particles in the Galactic halo \citep{dobler11a}.

The goal of this letter is not to delve into the origin question any
further, but rather to address one of the main discrepancies between
the microwave and gamma-ray haze/bubbles; namely, the observation
that \citep[as pointed out by][]{su10}, the gamma-ray emission appears
to have a sharp edge at latitudes $|b|\approx50\degree$, while the
microwaves fall off in intensity closer to
$|b|\approx35\degree$ \citep[see][]{dobler12}.  In \refsec{methods} I
describe the component separation methods used to uncover both the
microwave and gamma-ray haze/bubbles in WMAP and \emph{Fermi} data,
and in \refsec{results} I will show that, in fact, the
microwaves \emph{also} have a sharp edge at $b\approx-50\degree$ that
is spatially coincident with the edge in the \emph{Fermi} emission.
In \refsec{summary} I summarize this work and describe the important
implications for this microwave edge on the interpretation of the
Galactic haze/bubbles.

\begin{figure}[!ht]
  \centerline{
    \includegraphics[width=0.49\textwidth]{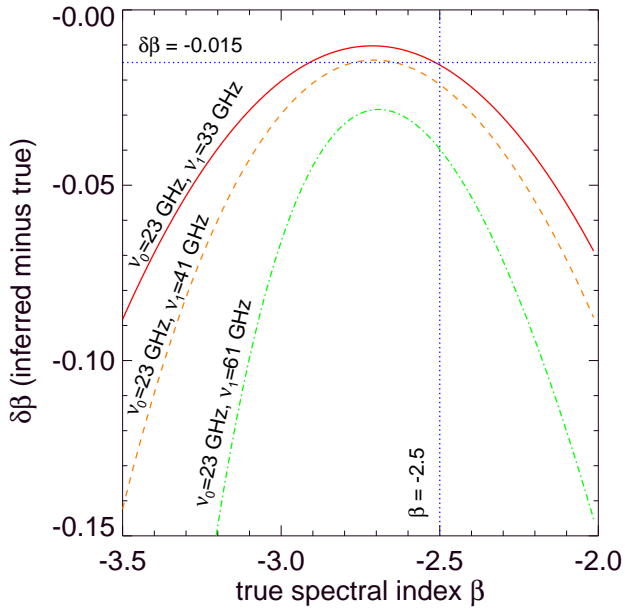}
  }
\caption{
  The difference between the inferred and true spectral index as a
  function of the true index for the CMB estimate used in this
  letter.  The inferred haze spectrum measured by \cite{dobler08a} and
  \cite{dobler12} is $\beta_H \approx -2.55$.  Since the CMB bias for
  a $T \propto \nu^{\beta}$ foreground with $\beta \approx -2.5$ is
  only 0.015, the measured spectrum is \emph{not} significantly biased.
  This implies that the electron population has an energy distribution
  of $dN/dE \propto E^{-2.1}$ at $E\sim10$ GeV.
}
\label{fig:cmb5_deltabeta}
\end{figure}

\begin{figure*}[!ht]
  \centerline{
    \includegraphics[width=0.98\textwidth]{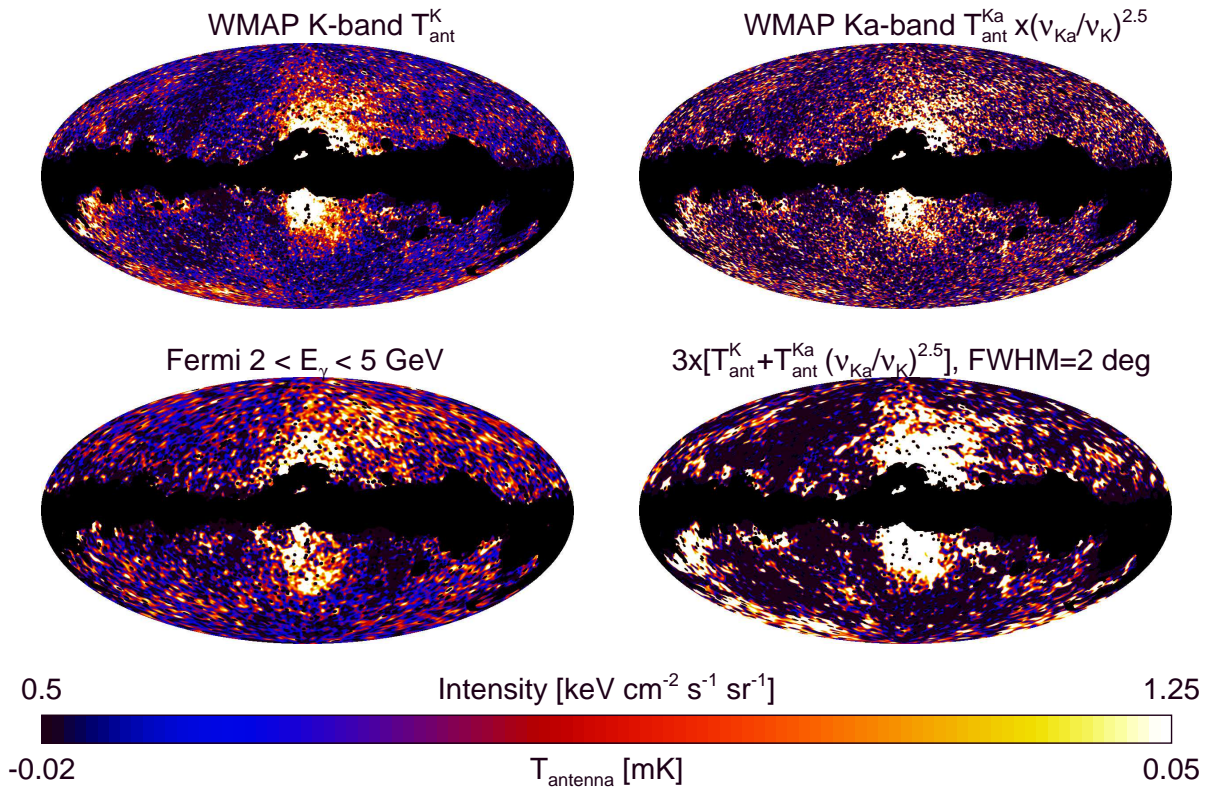}
  }
\caption{
  \emph{Top row:} the microwave haze at WMAP K- and Ka-band smoothed
  to $1\degree$.  As noted in \cite{dobler12}, the microwaves seem to
  fade quickly below $b \approx -35\degree$ in contrast to the
  \emph{Fermi} haze/bubbles (\emph{lower left}) which continue down to
  $b \approx -50\degree$.  The \emph{Fermi} residual is just the
  difference between the data and the \emph{Fermi} diffuse model for
  visualization (see \refsec{methods}).  However, when smoothing the
  WMAP data to $2\degree$ and stacking with weights given by
  $(\nu/\nu_{\rm K})^{2.5}$ it can be seen in the \emph{lower right}
  panel that there is an edge in the microwaves at $b \approx
  -50\degree$ as well.
}
\label{fig:wmap_fermi_stck}
\end{figure*}

\begin{figure*}[!ht]
  \centerline{
    \includegraphics[width=0.85\textwidth]{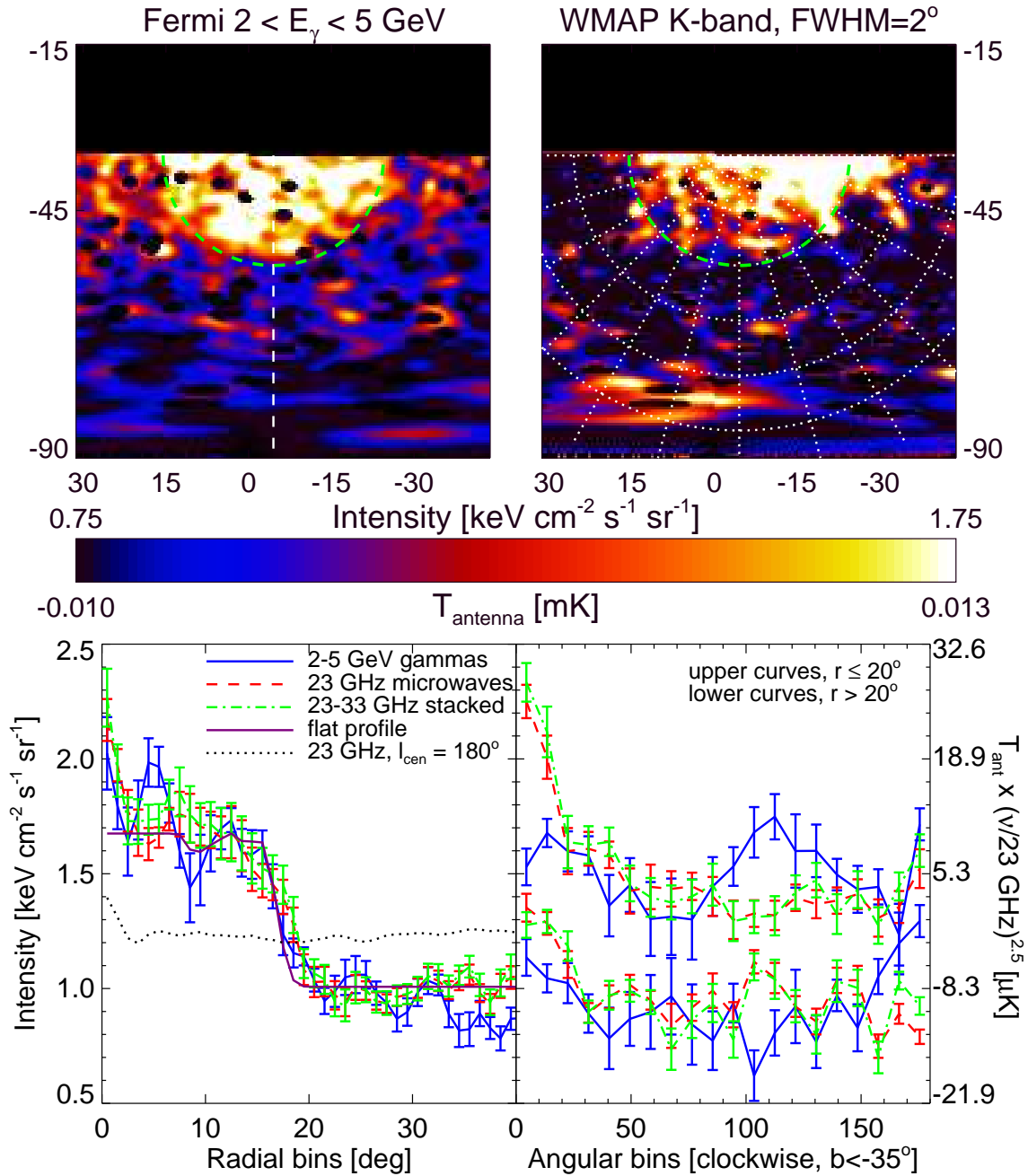}
  }
\caption{
  \emph{Top left:} the southern \emph{Fermi} bubble at latitudes $b <
  -35\degree$ where the edges are most discernible in the gammas.
  \emph{Top right:} the same region but for the WMAP K-band
  haze/bubbles smoothed to $2\degree$.  There is a clear spatial
  correspondence between the \emph{Fermi} haze/bubbles and the WMAP
  haze, \emph{including} an edge at high latitudes.  Taking the center
  of the emission to be $(\ell,b)_{\rm cen} =
  (-4.5\degree,-35.0\degree)$ (dashed white line) and plotting the
  intensity of the gammas and microwaves as a function of distance
  from the center (i.e., integrating over the angular bins shown in
  dotted lines in the \emph{upper right} panel) reveals an unambiguous
  detection of an edge at high latitudes that is spatially coincident
  with the edge in the \emph{Fermi} haze/bubbles as shown in the
  \emph{bottom left} panel.  This emission is well fit by a flat
  brightness profile with a sharp edge at $r\sim17\degree$ from the
  center (performing the same annular bin using $\ell_{\rm cen} =
  180\degree$ yields no such feature).  \emph{Lower right:} the same
  but integrating over radial bins (for two ranges in $r$) as a
  function of angular bin.
}
\label{fig:wmap_fermi_edge}
\end{figure*}

\begin{figure*}[!ht]
  \centerline{
    \includegraphics[width=0.85\textwidth]{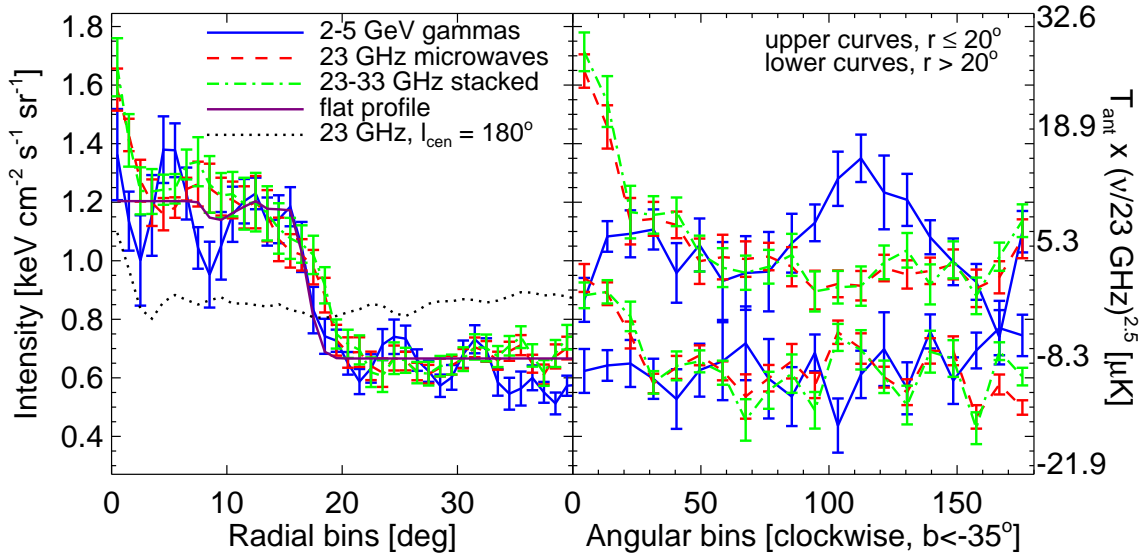}
  }
\caption{
  The same as the bottom panels of \reffig{wmap_fermi_edge}, but using
  the \emph{Fermi} diffuse model in the template fit.
}
\label{fig:wmap_fermi_edge_mod}
\end{figure*}

\begin{figure}[!ht]
  \centerline{
    \includegraphics[width=0.49\textwidth]{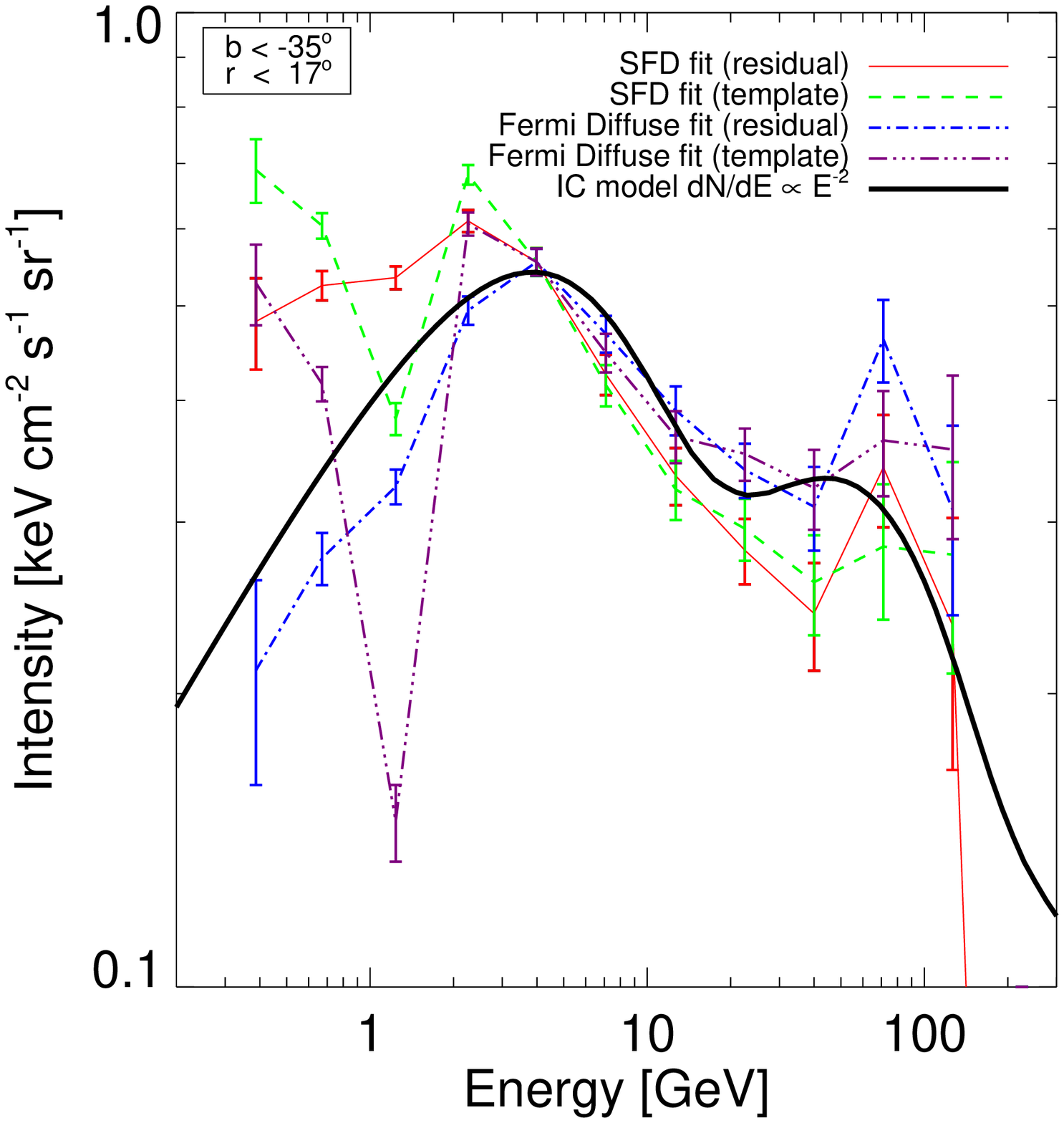}
  }
\caption{
  The spectrum of the \emph{Fermi} haze/bubble for $b<-35\degree$ and
  within $r<17\degree$ of $(\ell,b) = (-4.5\degree,-35\degree)$.  The
  four different lines represent spectra derived from haze template or
  residual amplitudes for both SFD and \emph{Fermi} diffuse model
  fits.  The spectra have been normalized to the SFD residual spectrum
  at 4 GeV.  While the uncertainty at low energies is large, the four
  spectral estimates agree above 4 GeV and are consistent with IC
  emission from an electron population with $dN/dE \propto E^{-2}$.
  The spectrum is softer than at lower latitudes (suggesting that the
  IC signal is due primarily to CMB scattering at distances $>6$
  kpc above the Galactic plane), which argues for a leptonic (IC)
  rather than hadronic ($\pi^0$ decay) origin.
}
\label{fig:fermi_cap_spec}
\end{figure}

\section{Methods}
\label{sec:methods}
  The most straightforward method for uncovering the Galactic
haze/bubbles in both the WMAP and \emph{Fermi} maps is via ``template
fitting'' which uses maps at other wavelengths to morphologically
trace known emission mechanisms in the data.  A simple linear
regression of these templates (using an appropriate mask) against the
data yields template amplitudes which can be used to ``peel away''
these foregrounds.  In this letter, the templates are fit at each
frequency (energy) for the WMAP (\emph{Fermi}) data implying that no
constraint is put on the shape of the spectrum of each emission
mechanism, though it is assumed that the spectrum does not vary
significantly with position.  The details of the template fitting used
here can be found in \cite{dobler08a} and \cite{dobler12} (hereafter
DF08 and D12 respectively) for the microwaves and \cite{dobler10} for
the gammas.

\subsection{WMAP analysis}

For the WMAP analysis, the templates used are the \cite{schlegel98}
(SFD) dust map evaluated at 94 GHz by \cite{finkbeiner99} (FDS),
the \cite{haslam82} 408 MHz map, and the \cite{finkbeiner03} H$\alpha$
composite map.  These templates are meant to trace the three primary
Galactic emission mechanisms at microwave wavelengths: thermal and
spinning dust (FDS), soft synchrotron (Haslam), and free--free
(H$\alpha$).  In addition, I include the haze and hard disk bivariate
Gaussian templates used in D12.  Because the CMB is of comparable
brightness to the haze at WMAP wavelengths, I presubtract the CMB5
estimate for the haze given by DF08.  Pixels for which the dust
extinction at H$\alpha$ is greater than 1 magnitude and for which the
H$\alpha$ intensity is greater than 10 Rayleigh are masked in the fit,
as well all point sources in the WMAP and \emph{Planck} ERCSC (30 GHz
to 143 GHz) catalogs.

These Galactic templates are fit to the WMAP data via the regression
equation ${\bf w}_{\nu} - {\bf c} = {\bf P}\vec{a}_{\nu}$ where ${\bf
w}_{\nu}$ is a map of the WMAP data at frequency $\nu$, ${\bf c}$ is
the CMB5 estimate, and ${\bf P}$ is a matrix of template maps.  The
equation is solved for the coefficients $\vec{a}_{\nu} = ({\bf P}^{\rm
T}{\bf n}_{\nu}^{-1}{\bf P})^{-1}({\bf P}^{\rm T}{\bf
n}_{\nu}^{-1})({\bf w}_{\nu}-{\bf c})$ where ${\bf n}$ is the mean
noise in a given band.  The haze residual in each band is defined as
$R_H \equiv {\bf w}_{\nu} - {\bf c} - {\bf P}\vec{a}_{\nu} +
a^{H}_{\nu} \times H$ where $H$ is the haze template.  That is, it is
the residual of the regression plus the amount of haze template
removed.  In order to account for spectral variations with position,
the fit is performed independently on the regions given in
D12.\footnote{An extreme example of this technique are pixel-by-pixel
fits of the data which use a combination of spectral and spatial
templates.  While the flexibility of these models makes the haze
analysis more difficult, \cite{pietrobon11} showed that the microwave
haze is indeed recoverable with these techniques.}  A composite $R_H$
is constructed from the union of those regions.

\subsection{\emph{Fermi} analysis}

At \emph{Fermi} energies, there are three main sources of diffuse
gamma-ray emission from the Galaxy: bremsstrahlung, IC, and $\pi^0$
decay.  The latter component is the dominant emission mechanism at
high latitudes.  For the purpose of comparing the haze/bubbles
in \emph{Fermi} to the emission in WMAP, this letter concentrates on
the extreme high Galactic southern latitudes, $b < -35\degree$.  High
northern Galactic latitudes are contaminated by dust-correlated
emission in both the microwaves \citep[likely spinning dust;
see][]{draine98,dobler09} and the gammas ($\pi^0$ decay) and make a
clear separation of the haze/bubbles emission difficult in both
datasets.

Because the $\pi^0$'s are created via collisions of cosmic-ray protons
with the ISM, I use the SFD map of dust column density as a template
for this emission since it is a reasonable tracer of dust and gas in
our Galaxy.  It is important to note that this template is an
integrated column density while $\pi^0$ emission is proportional to
the ISM density times the proton number density, and so line-of-sight
effects render the SFD map an imperfect tracer of $\pi^0$ emission.
This has dramatic effects at low latitudes in the haze/bubbles regions
as pointed out by \cite{dobler11a}, however for $b<-35\degree$ as is
used here, a simple two template \citep[SFD plus bivariate Gaussian;
see][]{dobler10} description of the data is sufficient to isolate
the \emph{Fermi} haze/bubbles emission.

As in \cite{dobler10} the fit minimizes the log-likelihood, $\ln
{\mathcal L} = \sum_i [k_i \ln\mu_i - \mu_i - \ln(k_i!)]$, where $k_i$
is the map of observed counts at pixel $i$, $\mu$ is a synthetic
counts map given by $\mu(E) = S(E) \times ({\rm mask}) \times ({\rm
exposure})$, and $S(E)$ is the synthetic sky map $S(E) = b_{\rm
SFD} \times {\rm SFD} + b_H \times H + b_{\rm uni}$.  This
log-likelihood is minimized over the template amplitudes $b_{\rm
SFD}$, $b_H$, and $b_{\rm uni}$ (a spatially uniform contribution) for
pixels outside of the mask and for $b < -35\degree$.\footnote{All
gamma-ray results in this letter are derived with
1.6-year \emph{Fermi} maps constructed as described
in \cite{dobler10}.}  I emphasize that this is not meant to represent
a perfect model for the diffuse emission from our Galaxy, but rather
serves to effectively isolate the \emph{Fermi} haze/bubbles emission
so that it can be morphologically compared to the microwave emission.
For a thorough analysis of Galactic diffuse emission observed
by \emph{Fermi} see \cite{porter12}.  Lastly, the analysis was also
redone using the \emph{Fermi} diffuse model
(\texttt{gll\_iem\_v02.fit}\footnote{
\scriptsize\url{http://fermi.gsfc.nasa.gov/ssc/data/access/lat/BackgroundModels.html}}) as opposed to the SFD map as a template and the results are not
significantly changed.

\section{Results}
\label{sec:results}
  Directly comparing the haze/bubbles residuals at high latitude will
involve weighted stacking of $R_H$ in multiple WMAP bands, and so it
is useful to estimate the extent to which the spectrum of the
microwave emission is affected by the CMB bias described in DF08.
This bias comes from the fact that any CMB estimate used
in \refsec{methods} will inherently have some residual foregrounds
after cleaning emission from the Galaxy, and since that CMB estimate
is presubtracted with a fixed CMB spectrum, this imprints a bias on
the \emph{inferred} haze spectrum (see DF08).  The CMB estimate used
here is generated by the linear combination of the WMAP data that
minimizes the variance in unmasked pixels of ${\bf c}
= \sum_{j} \zeta_j {\bf w}_j^{\prime}$ where ${\bf w}_{j}^{\prime}$ is
the WMAP data minus the FDS prediction for thermal dust emission in
each band $j$ and $\sum_{j} \zeta_{j} \equiv 1$ for ${\bf w}$ in
thermodynamic $\Delta T$ units.

This last constraint on $\zeta$ has the consequence that the measured
spectrum cannot be used to infer the true spectrum.  That is, while
the amplitude of the bias can be estimated as in DF08, the exact bias
cannot be known.  However, it is possible to answer the question: for
a foreground with a given true spectrum $T_{\rm
antenna} \propto \nu^{\beta_{\rm true}}$, what would be
the \emph{inferred} spectral index $\beta_{\rm infer}$?  This
$\delta\beta \equiv \beta_{\rm infer} - \beta_{\rm true}$ is shown as
a function of $\beta_{\rm true}$ in \reffig{cmb5_deltabeta}.  Despite
the very large possible CMB bias, for a true spectral index of
$\beta_{\rm true} = -2.5$, the inferred spectral index when comparing
K- to Ka-band would only be biased by $\delta\beta=-0.015$ given the
CMB5 coefficients $\zeta = (0.11, -0.48, 0.12, 0.16, 1.09)$.  The
implication is that the spectrum measured by DF08 and D12 for the
microwave haze $T_H \propto \nu^{\beta_H}$ with $\beta_H \approx -2.5$
is likely \emph{not significantly biased} and the underlying electron
population is very close to $dN/dE \propto E^{-2}$.

The full sky residuals $R_H^W$ and $R_H^F$ --- the haze/bubbles in
WMAP and \emph{Fermi} data --- are shown in \reffig{wmap_fermi_stck}
and illustrate several of the key characteristics noted in previous
studies: scaling $R_H^W$ by $\nu^{2.5}$ yields roughly equal
brightness indicating a $T \propto \nu^{-2.5}$
spectrum \citep{dobler08a}, $R_H^W$ and $R_H^F$ are roughly spatially
coincident at low latitudes \citep{dobler10,su10}, at $1\degree$
smoothing $R_H^W$ appears to fade quickly below
$b=-35\degree$ \citep{dobler12}, and $R_H^F$ appears to have a sharp
``edge'' at $|b|\sim50\degree$ \citep{su10}.  The lack of a similar
edge in the WMAP data at $b\sim-50\degree$ in WMAP has led to some
ambiguity about whether the WMAP haze and the \emph{Fermi}
haze/bubbles are in fact the same structure.  However, creating the
weighted stack of the WMAP emission $R_H^{\rm K} + (\nu_{\rm
Ka}/\nu_{\rm K})^{2.5}\times R_H^{\rm Ka}$ and smoothing to $2\degree$
(i.e., the same smoothing as the \emph{Fermi} map) reveals an edge in
the microwave haze that appears coincident with the southern edge in
the gamma-ray bubbles.

To assess the significance of this feature, I zoom in on the extreme
southern latitudes with $b<-35\degree$ in \reffig{wmap_fermi_edge}
(corresponding to heights $>6$ kpc above the GC), and a clear sharp
edge is evident in both the gamma-rays \emph{and the microwaves}.  As
noted in \cite{su10} the center of the \emph{Fermi} bubble at these
latitudes is roughly $\ell\sim-4.5\degree$.  Binning the sky into
polar bins centered on the bubble center $(\ell_{\rm cen},b_{\rm cen})
= (-4.5\degree,-35\degree)$, integrating annuli with unmasked pixel
latitudes $<-35\degree$ (see figure), and plotting as a function of
distance from the bubble center, the lower left panel
of \reffig{wmap_fermi_edge} shows the southern \emph{Fermi} bubble
edge described in \cite{su10}.  The same plot generated with the
microwaves also shows a clear edge that is \emph{spatially coincident
with the \emph{Fermi} bubble edge}, located at roughly $r = 17\degree$
from the bubble center.  For the $2\degree$ smoothing shown here, the
statistical significance of the edge identified in this way is high
for both the K-band microwave haze/bubbles and the K-, Ka-band
weighted stack (a null test calculated by setting $(\ell_{\rm
cen},b_{\rm cen}) = (-180\degree,-35\degree)$ shows no evidence for an
edge at the Galactic anti-center).  Finally, by integrating a uniform
brightness, infinitely sharp edge (smoothed to $2\degree$ and with the
same mask) in the same way, \reffig{wmap_fermi_edge} shows
that \emph{Fermi} and WMAP bubbles are consistent with a sharp edge.

While this is a clear detection of an edge in the WMAP haze, there is
angular dependence in the microwave emission that is not readily
apparent in the gamma-rays.  \reffig{wmap_fermi_edge} shows the
emission as a function of annulus angle both inside and outside the
haze/bubbles ($r<20\degree$ and $r>20\degree$ from $(\ell_{\rm cen},
b_{\rm cen})$) in the lower right panel and indicates an ``arm'' of
emission in the microwaves for annular angles less than
$\sim30\degree$.  No such arm is evident in the gamma-rays.  For
annular angles greater than $30\degree$ there is no other clear
structure present in the microwaves though the microwave bubble
emission is detected in each angular bin (i.e., there is an excess of
emission interior to the microwave edge compared to
exterior).  \reffig{wmap_fermi_edge_mod} shows the same results, but
using the \emph{Fermi} diffuse model template in the fit.  While the
overall amplitude of the gamma-ray signal is slightly lower, the edge
is still spatially coincidence with the edge in microwaves.

\reffig{fermi_cap_spec} shows the spectrum of the \emph{Fermi} 
haze/bubble for $b<-35\degree$ and within $r<17\degree$ of the bubble
center for four different spectral estimates: using the SFD map or
the \emph{Fermi} diffuse model in the template fit and estimating the
spectrum via unmasked pixels in $a_{\nu}^H \times H$ or $R_H$.  Below
$\sim4$ GeV, the four different types of spectral estimates differ,
indicating a systematic bias in the derived spectrum due to imperfect
template approximations and/or brighter, softer components leaking
into the residuals.  However, above $\sim4$ GeV, the spectra are all
consistent with IC from electrons scattering predominantly off of CMB
photons and having a spectrum $dN/dE \propto E^{-2}$ (i.e., the same
as that required to make the $T\propto\nu^{-2.5}$ microwave haze
spectrum).  It is interesting to note that this spectrum of the
``cap'' of the bubble is somewhat softer than that found
by \cite{dobler10} and \cite{su10} who include lower latitudes, in
agreement with the IC scenario (since the higher energy optical and IR
components of the ISRF have lower amplitude at high
latitudes)\footnote{For this calculation, the \texttt{GALPROP} ISRF
estimate was used (see \url{http://galprop.stanford.edu/}).  Since
this estimate only goes up to $\sim5$ kpc in height above the plane, I
take the optical and IR components to be 75\% of the 5 kpc value.}
and inconsistent with a hadronic scenario, which is also disfavored by
the detection of the microwave haze at $b<-35\degree$
in \reffig{wmap_fermi_edge}.

Since the IC emission is a function of the number density of electrons
$dN/dE$ and the ISRF intensity while the synchrotron intensity depends
on $dN/dE$ and the magnetic field $B$, an estimate of $B$ can be made
under several simplifying assumptions.  Given the (CMB dominated) ISRF
model above and assuming the same $dN/dE = N_0 \times E^{-2}$ for both
signals, a uniform magnetic field and volume emissivity model above 6
kpc ($b < -35\degree$) yields $B \sim 5$ $\mu$G within the southern
bubble.

\section{Summary}
\label{sec:summary}
  Using 7-year WMAP data to isolate the microwave haze and comparing
this to the \emph{Fermi} haze/bubbles at southern Galactic latitudes
less than $-35\degree$, I have presented the detection of a sharp edge
in the microwave haze that is coincident with the
\emph{Fermi} bubble edge.  This microwave bubble edge is evident 
when smoothing the haze to 2$\degree$ in both the K-band WMAP data as
well as a stack of K- and Ka-band weighted by $\nu^{2.5}$, where $\nu$
is the microwave band.  I have also shown explicitly that, for the CMB
estimate used in this study as well as \cite{dobler12}, the
$\nu^{-2.5}$ spectrum of the microwave haze/bubbles is not
significantly biased by systematics indicating that the electrons
responsible for generating this synchrotron have a number density as a
function of energy $dN/dE \propto E^{-2}$, in excellent agreement with
the inverse Compton interpretation of the gamma-rays.

The detection of an edge in the microwave haze $\sim 50\degree$ above
the plane and coincident with the \emph{Fermi} haze/bubbles has
several important consequences.  First, it proves conclusively that
the microwave and and the \emph{Fermi} bubbles are the same structure
observed at multiple wavelengths.  Second, given the vastly different
experiments, the detection of an edge in microwaves proves that the
edge in gammas is real and not due to processing artifacts or low
photon counts in the maps generated by \cite{dobler10}
and \cite{su10}.  Finally, the sharp edge, coupled with the hard
spectrum of the emission, suggests a transient event for the origin of
the microwave haze indicating that it is a separate component of
diffuse emission in our Galaxy and not merely a spatial variation in
the spectral index of the disk synchrotron.

\vskip 0.15in {\bf \noindent Acknowledgments:} I thank Peng Oh, 
Christoph Pfrommer, Krzysztof Gorsk\'{i}, Neal Weiner, and Doug
Finkbeiner for useful conversations.  This work has been supported by
the Harvey L.\ Karp Discovery Award.


\end{document}